\documentclass[twocolumn,superscriptaddress,floatfix,prb,amsmath,aps,showpacs]{revtex4}
\usepackage{graphicx}
\usepackage{bm}
\begin{document}
\bibliographystyle{apsrev}
\def\sa{\section}
\def\sb{\subsection}
\def\sc{\subsubsection}
\def\ind{\ \ \ \ }
\def\be{\begin{equation}}
\def\ee{\end{equation}}
\def\bea{\begin{eqnarray}}
\def\eea{\end{eqnarray}}
\def\ba{\begin{array}}
\def\ea{\end{array}}
\def\nn{\nonumber}
\def\ben{\begin{enumerate}}
\def\een{\end{enumerate}}
\def\fn{\footnote}
\def\rd{\partial}
\def\rot{\nabla\times}
\def\r{\right}
\def\l{\left}
\def\gt{\rightarrow}
\def\cf{\leftarrow}
\def\bw{\leftrightarrow}
\def\ra{\rangle}
\def\la{\langle}
\def\bla{\big\langle}
\def\bra{\big\rangle}
\def\Bla{\Big\langle}
\def\Bra{\Big\rangle}
\def\ddt{{d\over dt}}
\def\rdt{{\rd\over\rd t}}
\def\rdx{{\rd\over\rd x}}
\def\bb{\begin{thebibliography}{99}}
\def\eb{\end{thebibliography}}
\def\bit{\bibitem}
\def\bc{\begin{center}}
\def\ec{\end{center}}

\title{$SU(2)$ slave-boson formulation of spin nematic       
states in $S=\frac{1}{2}$ frustrated ferromagnets}
\author{Ryuichi Shindou}
\author{Tsutomu Momoi}
\affiliation{Condensed Matter Theory Laboratory,  RIKEN,
2-1 Hirosawa, Wako, Saitama 351-0198, Japan}
\begin{abstract}
{An $SU(2)$ slave boson formulation of bond-type spin nematic orders is
developed in frustrated ferromagnets,
where the spin nematic states are described as the
resonating {\it spin-triplet}
valence bond (RVB) states. 
The $\bm d$-vectors of spin-triplet pairing ansatzes play the role
of the directors in the bond-type spin quadrupolar states.
The low-energy excitations around such spin-triplet RVB ansatzes
generally comprise the (potentially massless) gauge bosons,
massless Goldstone bosons, and spinon individual excitations.
Extending the projective symmetry group argument
to the spin-triplet ansatzes, we show how to
identify the number of massless gauge bosons efficiently. 
Applying this formulation, we next (i) enumerate possible mean field
solutions for the $S=\frac{1}{2}$ {\it ferromagnetic} $J_1$-$J_2$ Heisenberg
model on the square lattice, with ferromagnetic nearest neighbor $J_1$
and competing antiferromagnetic next-nearest neighbor $J_2$, and
(ii) argue their stability against small gauge
fluctuations. As a result, two stable spin-triplet RVB ansatzes are found
in the intermediate coupling regime around $J_1 : J_2 \simeq 1:0.4$.
One is the $Z_2$ Balian-Werthamer
(BW) state stabilized by the Higgs mechanism
and the other is the $SU(2)$ chiral $p$-wave (Anderson-Brinkman-Morel)
state stabilized by the Chern-Simon mechanism. 
The former $Z_2$ BW state in fact shows
the same bond-type spin quadrupolar order as found in
the previous exact diagonalization study
[N.~Shannon {\it et al.}, 
Phys.\  Rev.\  Lett.\ {\bf 96}, 027213 (2006)].}
\end{abstract}
\maketitle
\section{introduction}

Recent theoretical progress has revealed that 
a certain class of frustrated 
magnets\cite{ag,Chubukov,cc,ldlst,sms,Vekua,HikiharaKMF, 
momoi,MomoiSS,MomoiKS} shows   
spin nematic states\cite{cl,ag} as their magnetic ground 
states,    
where the spin quadratic
tensor, $K_{jl,\mu\nu} \equiv
\langle S_{j\mu} S_{l\nu} \rangle - \frac{\delta_{\mu\nu}}{3}
\langle {\bm S}_{j}\cdot {\bm S}_{l}\rangle$ with $(\mu,\nu=1,2,3)$,
exhibits a long-range order, while the spin moment
$\langle S_{j\mu} \rangle$ remains disordered.
Such spin nematic states can 
be classified into the chiral type ($p$-nematic) and non-chiral 
type ($n$-nematic) states,\cite{ag} according to the parity of 
the spin quadratic tensor. Namely, the antisymmetric 
quadratic tensor $P_{jl,\lambda}\equiv 
\epsilon_{\lambda\mu\nu} K_{jl,\mu\nu}$ is nothing but the  
the vector chirality, while the
symmetric part --- non-chiral one --- plays the role of the 
spin quadrupolar moment, 
$Q_{jl,\mu\nu} \equiv  \frac{1}{2}(K_{jl,\mu\nu} + K_{jl,\nu\mu})$.
The latter ordered state is a spin analogue of the nematic state well
known
in liquid crystals,\cite{chlu} where the order parameter is
characterized by
the so-called `director vector' ${\bm d}({\bm r})$ in the form 
\begin{eqnarray}
Q_{\mu\nu}({\bm r}) = d_{\mu}({\bm r}) d_{\nu}({\bm r})
- \frac{1}{3} \delta_{\mu\nu} |{\bm d}({\bm r})|^2. \label{dir0}
\end{eqnarray}
From this analogy, the spin quadrupolar states are often dubbed
simply as the
`spin-nematic' states.\cite{ag} 


Depending on how the spin quadrupolar moments are 
microscopically organized, 
spin nematic states have two 
distinct classes; 
(i) site-type nematic states\cite{cl,vmm,np,ac,jmmsz,haradak,tsune,lmp} 
and (ii) bond-type nematic states.\cite{ag,Chubukov,cc,ldlst,sms,
Vekua,HikiharaKMF,momoi,MomoiSS,MomoiKS}  
The former types of nematic orders are realized in 
the spin 1 bilinear-biquadratic model,
$\mathcal{H}_{S=1}= \sum_{\langle ij\rangle}[J {\bm S}_i\cdot {\bm S}_j
+ K ({\bm S}_i\cdot {\bm S}_j)^2]$, where the 
quadrupolar moments constituted {\it at respective sites} 
exhibit the long-range order due to the strong 
biquadratic coupling.\cite{cl,vmm,ac,np,haradak}  
Ground state wavefunctions of these site-type nematic states  
can be essentially factorized into  
decoupled `vacuums', which are defined 
on respective sites. Thus, their spin-wave 
theories\cite{vmm,ac,np,jmmsz,tsune,lmp} including 
low-energy effective theories\cite{ik}  
were well-established. Namely, the elementary excitation   
around such a site-factorized vacuum is also given by a 
linear combination of bosons introduced at respective sites.


The simplest localized spin models which allow the 
second class of spin nematic states -- bond-type 
nematic states -- are the spin one half 
frustrated ferromagnets,\cite{ag,Chubukov,cc,ldlst,sms,Vekua,HikiharaKMF, 
momoi,MomoiSS,MomoiKS} which could be realized in a certain
family of layered cuprates\cite{kageyama,oba,tsujimoto,Drechsler1,Drechsler2}
and vanadates\cite{V_exp1,V_exp2} and also in solid $^3$He films.\cite{solid3He}
For example, in (CuX)LaNb$_2$O$_7$ (X=Cl,Br),\cite{kageyama,oba}
Cu$^{2+}$ ions, having a localized spin $\frac{1}{2}$,
compose a square lattice, while the anion X$^{-}$
locates at the center of the square
instead of the bond center.
As a result, the nearest neighbor (NN) 
exchange interaction $J_1$ between the 
localized spins becomes ferromagnetic
because of the Goodenough-Kanamori rule,\cite{kg}
while the next nearest neighbor (NNN) interaction $J_2$ 
becomes antiferromagnetic; the model-Hamiltonian is given by 
\begin{eqnarray}
\mathcal{H} = - J_1 \sum_{\langle j,l\rangle}
{\bm S}_j\cdot {\bm S}_l + J_2 \sum_{\langle\langle j,l\rangle\rangle}
{\bm S}_j\cdot {\bm S}_l,
\label{ff}
\end{eqnarray}
with $J_1,J_2 > 0$.
The preceding exact diagonalization (ED)
studies for this spin one half square lattice $J_1$-$J_2$ model\cite{sms}
indicated that the $d$-wave {\it bond-type} spin
nematic order develops in the intermediate 
parameter region, $J_1\simeq 2J_2$.   
Namely, strong ferromagnetic 
exchange interactions favor the spin-triplet 
valence bond formations between two neighboring spin one halves, 
while, simultaneously, these two spin one halves     
try to change their partners quantum-mechanically by way of   
the NNN antiferromagnetic exchange interactions.  
This leads to a kind of {\it resonating}  
spin-triplet valence bond state, 
where the quadrupolar moment organized  
{\it at each neighbor bond} exhibits the 
following antiferro-type configuration with the   
uniform amplitude;   
\begin{eqnarray}
&&\hspace{-0.9cm}
Q_{\langle j,j+\hat{x}\rangle,22} - 
Q_{\langle j,j+\hat{x}\rangle,11} = 
Q_{\langle j,j+\hat{y}\rangle,11} - 
Q_{\langle j,j+\hat{y}\rangle,22} > 0 
. \label{dwave}
\end{eqnarray}
Similar bond nematic order phases were 
also found in other frustrated ferromagnets, such as  
a zigzag spin chain\cite{Chubukov,Vekua,HikiharaKMF}
containing ferromagnetic $J_1$ and a triangular lattice 
multiple-spin exchange model.\cite{momoi,MomoiSS,MomoiKS}

In contrast to the site-type nematic states, however, 
when attempting to construct a mean-field description of these 
bond nematic states [as well as their spin wave theories], 
one could immediately reach a more fundamental question;  
how their ground state wavefunctions themselves 
should be described?  
Namely, since a single spin one half at each site is 
supposed to participate {\it equally} 
in the spin-triplet formations on its 
four ferromagnetic bonds  
[in the square lattice case], 
their ground state wavefunctions are no longer  
described by any kind of `site-factorized wavefunctions'.  

In this paper, we will
construct an $SU(2)$ slave-boson mean-field theory of the
bond-type spin nematic states, which
are described as the resonating valence bond (RVB) states of
the {\it spin-triplet} bonds.
After splitting the original spin
operator into the bilinear of the spinon fields
(fermions),\cite{pwa,kma,shf,wen,fradkin}
$S_{j\mu} \equiv \frac{1}{2}
f^{\dagger}_{j\alpha}[\sigma_{\mu}]_{\alpha\beta}f_{j\beta}$,
we first introduce the spin-triplet pairing ansatzes into the
ferromagnetic exchange bonds as
\begin{align}
E_{ij,\mu} &\equiv \langle f^{\dagger}_{i\alpha}
[\sigma_{\mu}]_{\alpha\beta}
f_{j\beta} \rangle, \label{tri3-2} \\
D_{ij,\mu} &\equiv \langle f_{i\alpha}
[i\sigma_2\sigma_{\mu}]_{\alpha\beta} f_{j\beta} \rangle,
\label{tri3-1}
\end{align}
where ${\bm D}_{ij}$ (${\bm E}_{ij}$) describes
the ${\bm d}$-vector of the spin-triplet pair 
condensation\cite{su} (`spin-orbit' hopping integral) .
In fact, these two-types of  the
${\bm d}$-vectors, i.e. that 
in the particle-hole channel and in
the particle-particle channel, precisely mimic
the director vector ${\bm d}({\bm r})$ of 
nematic states in liquid crystals [see Eq.~(\ref{dir0})]; in the mean-field
approximation, the quadrupolar order parameter is given by
\begin{align}
Q_{jl,\mu\nu}=& -\frac{1}{2} \big(
E_{jl,\mu}E^{*}_{jl,\nu} - \frac{1}{3}\delta_{\mu\nu}
|{\bm E}_{jl}|^2 \big) \ +\ {\rm h.c.} \nn \\
&- \frac{1}{2} \big( D_{jl,\mu}D^{*}_{jl,\nu}
- \frac{1}{3}\delta_{\mu\nu}
|{\bm D}_{jl}|^2\big)\ +\ {\rm h.c.}. \label{dir}
\end{align}
Moreover, the vector chiral order parameter is
given by the products between these
two ${\bm d}$-vectors and their respective
spin-singlet ansatzes in the form\cite{ran}
\begin{align}
P_{jl,\lambda}=&
\frac{i}{2} \big(\chi_{jl} E^{*}_{jl,\lambda} -
\chi^{*}_{jl} E_{jl,\lambda} \big) \nn \\
&
- \frac{i}{2} \big(\eta_{jl} D^{*}_{jl,\lambda}
- \eta^{*}_{jl} D_{jl,\lambda}\big), \label{vc}
\end{align}
where $\chi_{jl}$ ($\eta_{jl}$) stands for
the spinless hopping integral (spin-singlet
pair condensation),\cite{pwa,kma,shf}
\begin{eqnarray}
\chi_{jl} \equiv
\langle f^{\dagger}_{j\alpha}f_{l\alpha}\rangle,
\ \  \eta_{jl} \equiv \langle f_{j\alpha}
[(-i)\sigma_2]_{\alpha\beta}f_{l\beta} \rangle. \label{sin3-12}
\end{eqnarray}
Thus, one can naturally employ the spin-triplet
slave boson theory as a mean-field description
of the spin nematic orders.

In Section\ II, we will introduce
an $SU(2)$-formulation of
the spin-triplet mean-field ansatzes, where
we extensively use the $2\times 2$ matrix representation
originally introduced by Affleck {\it et al.},\cite{IA}
instead of the usual Nambu vector.
This representation [see Eqs.~(\ref{AZHA}) and (\ref{tri2})]
clearly dictates that the low-energy
excitation around any spin-triplet RVB state generally
consists of (gapless) Goldstone boson and
(potentially gapless) gauge boson. It is widely known
that the existence of the gapless gauge fluctuations is
crucial to the instability of the starting
mean-field ansatzes.\cite{polyakov,wen,fradkin}
Thus, we will next argue the 
spin-triplet extension of the projective
symmetry group (PSG) arguments. Without resorting to
any microscopic
calculations, this extension enables us to identify
the number of the massless gauge
bosons for any given mixed ansatz having both
spin-triplet and spin-singlet link variables.

Armed with these general formulations,
we study in Section~III the
ferromagnetic $J_1$-$J_2$ Heisenberg square lattice model
defined in Eq.~(\ref{ff}), thereby finding two stable
spin-triplet RVB ansatzes
in the intermediate coupling region, $J_1 \simeq 2J_2$.
One is the Balian-Werthamer (BW) type triplet
pairing state\cite{BW}
having the coplanar configurations of the $\bm d$-vector,
$\hat{d}(k)\propto \hat{x}k_x + \hat{y}k_y$, while
the other is the chiral $p$-wave state\cite{ABM}
having its $\bm d$-vector all pointing in the same direction
$\hat{d}(k)\propto \hat{z}(k_x + ik_y)$
[see Fig.~\ref{mfsum}(b)].
The PSG arguments indicate that, in general, all the
non-magnetic (gauge) excitations in the BW state
have finite Higgs mass. Thus, this ansatz --- $Z_2$ BW state --- is
stable against any type of small gauge fluctuations.
On the other hand, the chiral
$p$-wave state does not break any of the $SU(2)$
gauge symmetry. 
Instead, it 
breaks the time-reversal symmetry and all the mirror
symmetries. As a result, nonmagnetic (gauge) bosons are
endowed by the Chern-Simon term 
with the topologically-induced mass. 
Thus, this $SU(2)$ chiral $p$-wave state is
also stable against any small gauge fluctuation.
Though both the BW and chiral $p$-wave states   
exhibit spin quadrupolar orders,
the BW state especially 
shows the same configuration of quadrupolar moments 
as the bond-type spin-nematic order found
in Ref.\ \onlinecite{sms}. Hence, we further
discuss possible experimental 
features of this BW state, mainly focusing 
on its magnetic excitations.

Section~IV is devoted 
to the summary and open issues. 
The relation between our $Z_2$ BW state and
the time-reversal topological insulator recently
discussed in the various literatures\cite{km1,z2qsh,fkm,QHZ,SRFL}
is briefly mentioned.
We also propose    
those combinations of the triplet and singlet
ansatzes which describe the vector chiral order
having {\it no} finite director vector,\cite{ldlst} i.e.
$P_{jl,\mu}\ne 0$ and $Q_{jl,\mu\nu} = 0$. 
Those readers who want to make  
the $SU(2)$ slave-boson study in frustrated 
ferromagnets to be a {\it controlled} analysis might as well  
consult the appendix~A, where we describe 
the large-$N$ generalization of frustrated 
ferromagnetic spin models.  

\section{$SU(2)$-formulation of spin-triplet RVB state}
\subsection{Matrix representation}
The slave-boson formulation begins with describing the
spin operator by the bilinear of fermion fields;
$2 S_{j\mu} \equiv
f^{\dagger}_{j\alpha}[\sigma_{\mu}]_{\alpha\beta}f_{j\beta}$.
The enlarged (fermion's) Hilbert space reduces to the
physical (spin's) Hilbert space,
provided that the following local constraints
are strictly observed at each site:
\begin{eqnarray}
f^{\dagger}_{j\alpha}f_{j\alpha}= 1, \ \ \
f^{\dagger}_{j\uparrow}
f^{\dagger}_{j\downarrow}= f_{j\downarrow}
f_{j\uparrow} = 0. \nn
\end{eqnarray}
In the partition function, these local constraints are
implemented as the coupling between the fermion (spinon)
fields and the temporal $SU(2)$ gauge fields
$a^{\nu}_{j,\tau}$ ($\nu=1,2,3$),\cite{pwa,kma,wen,fradkin}
\begin{align}
&
Z \equiv \int d\vec{a}_{\tau}
d\Psi^{\dagger} d\Psi \exp \bigg[- \int_0^{\beta}
d\tau \mathcal{L} \bigg],  \label{gauss} \\
&
\mathcal{L}\equiv \frac{1}{2} \sum_{j} {\rm Tr}\big[\Psi^{\dagger}_j
\big(\partial_{\tau} \sigma_0
 + \sum_{\nu=1}^3 ia^{\nu}_{j,\tau} \sigma_{\nu}\big)
\Psi_j\big] + \mathcal{H}, \nn
\end{align}
where $\Psi_{j}$ and $\Psi^{\dagger}_{j}$ stand for
the $2\times 2$ matrices
\begin{eqnarray}
\Psi_{j} \equiv
\left[\begin{array}{cc}
f_{j,\uparrow} & f_{j,\downarrow} \\
f^{\dagger}_{j,\downarrow} & - f^{\dagger}_{j,\uparrow} \\
\end{array}\right], \ \
\Psi^{\dagger}_{j} \equiv
\left[\begin{array}{cc}
f^{\dagger}_{j,\uparrow} & f_{j,\downarrow} \\
f^{\dagger}_{j,\downarrow} & - f_{j,\uparrow} \\
\end{array}\right].
\label{AZHA}
\end{eqnarray}
The spin Hamiltonian part $\mathcal{H}$
becomes quartic in the fermion field ($\Psi$-field).
Depending on the sign of the exchange interaction,
we decompose this quartic term into the
Stratonovich-Hubbard variables in two
alternative ways;
\begin{align}
Z =& \int dU^{\rm sin} dU^{\rm tri}
 d\vec{a}_{\tau}
d\Psi^{\dagger} d\Psi \exp
\bigg[- \int^{\beta}_0 d\tau \mathcal{L}\bigg], \label{gauss1} \\
\mathcal{L} =& \frac{1}{2} \sum_{j} {\rm Tr}\big[\Psi^{\dagger}_j
\big(\partial_{\tau}\sigma_0  + \sum_{\nu=1}^3 ia^{\nu}_{j,\tau}
\sigma_{\nu}\big)\Psi_j\big] \nn \\
&\hspace{-.3cm}
-  \frac{J_{1}}{4} \sum_{\langle jl \rangle}
\Big\{\big(-|{\bm E}_{jl}|^2 - |{\bm D}_{jl}|^2\big) + {\rm Tr}
\big[\Psi^{\dagger}_j
U^{\rm tri}_{jl,\mu}\Psi_{l}\sigma^{T}_{\mu}\big]
\Big\} \nn \\
&\hspace{-.3cm}
-  \frac{J_{2}}{4}  \sum_{
\langle\langle  jl \rangle\rangle}
\Big\{ \big(-|\chi_{jl}|^2 - |\eta_{jl}|^2\big)  + {\rm Tr}
\big[\Psi^{\dagger}_j
U^{\rm sin}_{jl}\Psi_{l}\big]\Big\}.  \nn \\
\label{tri2}
\end{align}
Namely, the triplet and singlet link-variables,
\begin{eqnarray}
U^{\rm sin}_{ij} \equiv
\left[\begin{array}{cc}
\chi^{*}_{ij} & \eta_{ij} \\
\eta^{*}_{ij} & - \chi_{ij} \\
\end{array}\right], \ \
U^{\rm tri}_{ij,\mu} \equiv
\left[\begin{array}{cc}
E^{*}_{ij,\mu} & D_{ij,\mu} \\
- D^{*}_{ij,\mu} & E_{ij,\mu} \\
\end{array}\right], \label{tri3}
\end{eqnarray}
are introduced as the auxiliary fields for the 
ferro- and antiferro-magnetic bonds, respectively.
This is simply because the sign of the ferromagnetic
exchange interaction generally allows us to perform
the gaussian-integration only over
the ${\bm d}$-vectors in the excitonic/Cooper channel.
In fact, this integration precisely reproduces the ferromagnetic
exchange interaction,
\begin{eqnarray}
- 4 {\bm S}_j \cdot {\bm S}_l
&=&  - \sum_{\mu=1}^{3}
(f^{\dagger}_{j\alpha}[\sigma_{\mu}]_{\alpha\beta}f_{l\beta})
(f^{\dagger}_{l\gamma}[\sigma_{\mu}]_{\gamma\delta}f_{j\delta})  \nn \\
&&
- \sum_{\mu=1}^3
(f^{\dagger}_{j\alpha}[\sigma_2\sigma_{\mu}]_{\alpha\beta}
f^{\dagger}_{l\beta})(f_{l\gamma}[\sigma_{\mu}\sigma_2]_{\gamma\delta}f_{j\delta}
), \nn
\end{eqnarray}
while that over the singlet variable leads
to the antiferromagnetic exchange
interaction,\cite{pwa,kma,shf,wen,fradkin}
\begin{eqnarray}
4{\bm S}_j \cdot {\bm S}_l &=& -
(f^{\dagger}_{j\alpha}f_{l\alpha})(f^{\dagger}_{l\beta}f_{j\beta})
\nn \\
&& -
(f^{\dagger}_{j\alpha}[\sigma_2]_{\alpha\beta}f^{\dagger}_{l\beta}
)(f_{l\gamma}[\sigma_2]_{\gamma\delta}f_{j\delta}). \nn
\end{eqnarray}
Thus, the slave-boson formulation of
{\it mixed} Heisenberg magnets generally
requires us to use the spin-triplet
link-variable $U^{\rm tri}_{jl,\mu}$ for
every ferromagnetic bond and the
spin-singlet link variable $U^{\rm sin}_{jl}$ for
every antiferromagnetic bond.

The saddle point solutions of Eq.~(\ref{tri2})
lead to the coupled gap equations for these
link-variables, i.e. Eqs.~(\ref{tri3-2}), (\ref{tri3-1}), and (\ref{sin3-12}),
whose right hand sides are self-consistently given
by these mean-fields themselves.
In terms of $U^{\rm tri}_{jl,\mu}$ and $U^{\rm sin}_{jl}$
thus determined, the spin quadrupolar moment and 
vector chirality 
are given by
\begin{align}
-2Q_{jl,\mu\nu} &= {\rm Tr}
[U^{\rm tri}_{lj,\mu}U^{\rm tri}_{jl,\nu}] -
\frac{\delta_{\mu\nu}}{3}\sum_{\lambda=1}^3{\rm Tr}
[U^{\rm tri}_{lj,\lambda}U^{\rm tri}_{jl,\lambda}], \label{nc} \\
-2i P_{jl,\lambda} &= {\rm Tr}
[U^{\rm sin}_{lj}U^{\rm tri}_{jl,\lambda}]. \label{c}
\end{align}
Comparing Eq.~(\ref{tri2}) with Eq.~(\ref{c}),
notice that the present $J_1$-$J_2$ model can
have spin quadrupolar order on ferromagnetic bonds,
but cannot have vector chirality on any links, since
$U^{\rm sin}_{lj}U^{\rm tri}_{jl,\lambda}=0$. 
Within our formalism, a naive mean-field description
of vector chiral orders becomes possible only in
those spin models having either symmetric
anisotropic exchange interactions or
antisymmetric anisotropic one.
In the next section, without making any
distinction between the $n$-nematic states and 
$p$-nematic ones, we will widely call 
those mean-field ansatzes 
having both finite triplet ansatz and singlet
ansatz as spin-triplet RVB states.

\subsection{Low-energy excitations
around spin-triplet RVB states}
To see the low-energy excitations around the
spin-triplet RVB ansatzes, let us first express
the spin operator in terms of the
$2\times 2$ matrix representation,\cite{IA}  $S_{j\mu} \equiv
\frac{1}{4}{\rm Tr}[\Psi^{\dagger}_{j}
\Psi_j \sigma^{T}_{\mu}]$. Namely, a spin rotation
is described by an $SU(2)$ matrix, say $h_j$,
applied from the left (right) hand side of
$\Psi^{\dagger}_j$
($\Psi_j$),
\begin{eqnarray}
\Psi_j \rightarrow \Psi_j h^{T}_j,
\ \ \Psi^{\dagger}_j \rightarrow h^{*}_j
\Psi^{\dagger}_{j}, \nn
\end{eqnarray}
while physical quantities
are invariant under any {\it local} $SU(2)$
gauge transformation applied from the right
(left) hand side of $\Psi^{\dagger}_j$ ($\Psi_j$):
\begin{align}
&\Psi_j \rightarrow g_j \Psi_j,\ \ \ \
\Psi^{\dagger}_j \rightarrow
\Psi^{\dagger}_{j} g^{\dagger}_{j}, \nn\\
&\big\{
U^{\rm tri}_{jl,\mu},U^{\rm sin}_{jl}
\big\}
\rightarrow
g_j \big\{
U^{\rm tri}_{jl,\mu},U^{\rm sin}_{jl}
\big\} g_l^\dagger. \nn
\end{align}
For example, both parts of the spin quadratic tensor, 
Eqs.~(\ref{nc}) and (\ref{c}), are invariant
under this local $SU(2)$ gauge transformation. 
In regard to these two symmetries,
any spin-triplet mean-field
ansatz is generally accompanied by
two types of low-energy excitations: the magnetic
ones (Goldstone bosons)\cite{nambu-goldstone} and the the non-magnetic
ones (gauge bosons).\cite{IA,pwa,kma,wen2,wen,fradkin}

The former excitations are semiclassically 
described by the deformations of 
the ${\bm d}$-vectors around its mean-field configuration,
\begin{eqnarray}
U^{\rm tri}_{jl,\mu} \equiv
\sum_{\nu=1}^3\bar{U}^{\rm tri}_{jl,\nu}
R_{\nu\mu}\Big(\frac{j+l}{2},\tau\Big) \label{rotat}
\end{eqnarray}
for $\mu=1,2,3$ with the 
$3 \times 3$ rotational matrix $\hat{R}(x,\tau)$. 
Such deformations cost infinitesimally small
energy in spin models with spin continuous symmetry,
provided that the variation of the rotation
is sufficiently slow in space and time.
This type of deformations describe the
Goldstone modes accompanying
the spontaneous symmetry breaking.

In addition to this conventional excitation,
a certain non-magnetic (gauge)
excitations also become massless, when
our starting mean-field ansatz is invariant
under a {\it continuous} gauge symmetry.\cite{pwa,kma,wen,fradkin} 
For example, assume that the 
invariant gauge group (IGG) contains the
$U(1)$ gauge symmetry
$\{e^{i\theta\sigma_3}|\theta \in [0,2\pi)\}$.
Namely, our mean-field ansatz is invariant under any
rotation around the $3$-axis in the gauge space,
\begin{eqnarray}
e^{i\theta \sigma_3}\big\{
\bar{U}^{\rm tri}_{jl,\mu},\bar{U}^{\rm sin}_{jl}
\big\} e^{-i\theta \sigma_3}  = \big\{
\bar{U}^{\rm tri}_{jl,\mu},
\bar{U}^{\rm sin}_{jl}\big\} \label{U1}
\end{eqnarray}
for $\mu=1,2,3$ 
and $\bar{a}^{\nu}_{j,\tau}=\delta_{\nu3}\bar{a}^{3}_{j,\tau}$.
Then, we can argue that
the following non-magnetic deformation also
comprises the gapless excitation: 
\begin{align}
\big\{
U^{\rm tri}_{jl,\mu},U^{\rm sin}_{jl}
\big\} &\equiv
\big\{\bar{U}^{\rm tri}_{jl,\mu} ,
\bar{U}^{\rm sin}_{jl} \big\}e^{ia_{jl}\sigma_{3}}, \label{s1} \\
a^{3}_{j,\tau} &\equiv  \bar{a}^{3}_{j,\tau}
+ a_{0}(j,\tau)  , \label{t1}
\end{align}
where $a_{jl}$ relates
to the spatial components of ``gauge fluctuations"
$a_{\alpha}(j,\tau)$ $(\alpha=1,\cdots,d)$ in the form
\begin{eqnarray}
a_{jl}(\tau) = (j-l)_{\alpha}a_{\alpha}(j,\tau). \label{t1-1}
\end{eqnarray}
Specifically, one can expand the effective action 
in terms of these variations 
$a_\alpha(j,\tau)$ ($\alpha=0,1,\cdots,d$), assuming
these fluctuations to be much smaller
than their units, $a_\alpha(j,\tau) \ll 2\pi$.
Up to their quadratic order, the effective action
generally reads as follows:
\begin{align}
F_{\rm gauge} &= \sum_{\alpha,\beta=0}^{d} \sum_{Q}
M_{\alpha\beta}(Q)  a_{\alpha} (Q) a_{\beta}(-Q) + \cdots,
\label{free} \\
a_{\alpha}(Q)
&= \frac{1}{\sqrt{N\beta}} \sum_{i\omega_n} \sum_{q}
e^{iqj -i\omega_{n}\tau} a_{\alpha} (j,\tau) \label{ft}
\end{align}
with $Q = (q,i\omega_m)$.
Then, taking into account the $U(1)$ gauge symmetry
of the mean-field ansatz,
one can specify the form of the $(d+1)\times (d+1)$ matrix
$\hat{M}(Q)$, such that the quadratic part in Eq.~(\ref{free})
reduces to the $U(1)$ gauge invariant form as in Eq.~(\ref{mw}).

To see this, introduce the following local
$U(1)$ gauge transformation in Eq.~(\ref{tri2}):
\begin{eqnarray}
\Psi^{\dagger}_{j}(\tau) &\rightarrow&
\Psi^{\dagger}_{j}(\tau)\ e^{i\theta_{j}(\tau) \sigma_3}, \nn \\
\Psi_{j}(\tau) &\rightarrow&
e^{-i\theta_{j}(\tau) \sigma_3}\Psi_{j}(\tau),  \nn
\end{eqnarray}
where $\theta_{j}(\tau)$ varies slowly
in space and time.
Under this transformation,
all changes in the link variables (\ref{s1}) are put into
the transformation, $a_{jl}\rightarrow a_{jl}+\theta_l-\theta_j$ 
and $a_{0} \rightarrow a_0 + \partial_{\tau}\theta$,  
due to the $U(1)$ symmetry in IGG. Thus 
the effective action around $Q \simeq 0$ is 
literally transformed as
\begin{eqnarray}
&&\hspace{-0.2cm}
F_{\rm gauge} \rightarrow \nn \\
&&\hspace{0.4cm}
\sum_{\alpha,\beta}\sum_{Q \simeq 0}
M_{\alpha\beta}(Q)
(a_{\alpha} + \partial_{\alpha}\theta)(Q)
(a_{\beta} + \partial_{\beta}\theta)(-Q).\nn
\end{eqnarray}
However, the
free energy should have been invariant under any gauge
transformation, since gauge degrees of freedom
can be absorbed into the integral variables,
$\Psi$-fields. This requires that
$\hat{M}(Q)$ must precisely reduce to zero
at $Q= 0$, so that
the quadratic part of the action takes the
$U(1)$ gauge invariant form, e.g.
\begin{eqnarray}
F^{U(1)}_{\rm gauge} = \frac{1}{8\pi}\sum_{Q \simeq 0}
\sum_{\alpha=0}^d
\frac{1}{{g}^2_{\alpha}}
f_{\alpha}(Q) f_{\alpha}(-Q) + \cdots,
\label{mw}
\end{eqnarray}
where $f_{\gamma}
\equiv \epsilon_{\alpha\beta\gamma}
\partial_{\alpha}a_{\beta}$ stands for the field
strength.\cite{pwa,kma,wen,fradkin}
It is well-known that this maxwell form 
does not suppress the gauge fluctuation efficiently. 
Especially, when the
mean-field ansatz have its fermionic excitations
fully gapped and when $d=2$, these massless 
gauge fluctuations destroy the mean-field ansatz
itself,\cite{polyakov,wen,fradkin}
apart from some exceptional 
cases.\cite{CS,Pisarski,wen3,khPW,volovik-yakovenko, diamantini}
Following the literature,\cite{wen} we call in this paper
such spin-triplet mean-field ansatz as
the gapped $U(1)$ [or $SU(2)$] state.

On the other hand,
if the starting mean-field
ansatz has {\it no} continuous invariant
gauge group (IGG) like in Eq.~(\ref{U1}), the local
minimum condition imposed on
mean-field ansatzes generally requires all
the eigenvalues of $\hat{M}(Q)$ to be positive.
Therefore, all the gauge fields have finite
Higgs mass around any $Q$;
\begin{eqnarray}
F^{Z_2}_{\rm gauge} = \sum_Q \sum_{\alpha=0}^d
\tilde{M}_{\alpha}(Q)
\tilde{a}_{\alpha}(Q) \tilde{a}_{\alpha}(-Q)
+ \cdots\label{Higgs}
\end{eqnarray}
with $\tilde{M}_{\alpha}(Q)>0$.
In contrast to the maxwell form discussed above, 
this finite Higgs mass suppresses any small gauge
fluctuation completely, 
so that the starting mean-field ansatz is always
guaranteed to be (at least locally) stable. Such
ansatzes are usually dubbed as the $Z_2$ state.

The efficient 
way to confirm 
the absence of the continuous IGG was
introduced by Wen,\cite{wen,wen2}
where he pointed out the sufficient condition for its absence.
We can extend his argument to the spin-triplet RVB states also.
To see this, let us begin with the calculation
of the $SU(2)$ flux defined on a plaquette by multiplying
link-variables along the closed loop
in a regular sequence,  where either $\bar{U}^{\rm sin}_{ij}$ or 
one of $\bar{U}^{\rm tri}_{ij,\mu}$ should be chosen on each link. 
For example, when the loop is given
by a triangular path
$i$~$\rightarrow$~$j$~$\rightarrow$~$k$~$\rightarrow$~$i$,
one can have an
$SU(2)$ flux by $\bar{U}_{ij}\bar{U}_{jk}\bar{U}_{ki}$,
which always transforms in a gauge-covariant way;
\begin{eqnarray}
\bar{U}_{ij}\bar{U}_{jk}\bar{U}_{ki}
\rightarrow g^{\dagger}_i \cdot
\bar{U}_{ij}\bar{U}_{jk}\bar{U}_{ki}
\cdot g_i \nn
\end{eqnarray}
under $\Psi_{j} \rightarrow g_j \Psi_j$.
As such, the {\it relative angle} subtended by
two distinct $SU(2)$ fluxes
derived from the same base-site, such as
$\bar{U}_{ij}\bar{U}_{jk}\bar{U}_{ki}$
and $\bar{U}_{ij}\bar{U}_{jl}\bar{U}_{li}$,
contains non-trivial {\it gauge-independent} information,
provided that the two triangular paths, $\langle ijk(i)\rangle$ and
$\langle ijl(i)\rangle$, are different with each other.
Note that, even out of the {\it same} triangular loop, we can have
{\it two} distinct fluxes, when one of its three links has
two different types of spin-triplet ansatzes,
$\bar{U}^{\rm tri}_{ij,1} \ne \bar{U}^{\rm tri}_{ij,2}$.
In this case,
we should regard that
$\bar{U}^{\rm tri}_{ij,1}\bar{U}^{\cdots}_{jk}
\bar{U}^{\cdots}_{ki}$ and
$\bar{U}^{\rm tri}_{ij,2}\bar{U}^{\cdots}_{jk}
\bar{U}^{\cdots}_{ki}$ are two
distinct fluxes obtained from the same base-site $i$.

Having all $SU(2)$ fluxes thus obtained
in hand, 
one can readily see that, {\bf (i)} if two distinct $SU(2)$ fluxes
obtained from the same base-site are not collinear with each other,
there is no continuous IGG in that mean-field ansatz.
{\bf (ii)} If all the distinct fluxes obtained
from the same base-site are pointed along one direction in the gauge
space, say along the $3$-axis, the ansatz could  
have a certain $U(1)$ gauge symmetry around this
$3$-axis, just like in Eq.~(\ref{U1}).  One can also confirm 
that, {\bf (iii)} the ansatz can be invariant under a certain
$SU(2)$ gauge symmetry [so-called $SU(2)$ state],
if all the $SU(2)$ fluxes are proportional to the unit matrix.

This `non-collinearity' argument 
of the $SU(2)$ fluxes concludes the (local) stability of
each ansatz against gauge fluctuations very efficiently, without
resorting to any microscopic calculation.
Thus, it substantially helps us to find a better
spin-triplet mean-field ansatz as in the case of
spin-single RVB ansatzes.\cite{wen,wen2}

\section{$J_1$-$J_{2}$ frustrated ferromagnetic square lattice
Heisenberg model}
In this section, we will apply the spin-triplet
slave-boson mean-field formulation onto
the spin-$\frac{1}{2}$ $J_1$-$J_2$
mixed Heisenberg model (\ref{ff}) on the square lattice
with ferromagnetic nearest neighbor (NN) $J_1$ and
antiferromagnetic next nearest neighbor (NNN) $J_2$.
As was described in the previous section, we
always decompose the ferromagnetic NN bond into the
spin-triplet ansatz and the antiferromagnetic
NNN bond into the spin-singlet ansatz.

\subsection{Mean-field solutions}
To be specific, we have numerically studied the various
local `stable' minima of the mean-field free energy
given in Eq.~(\ref{tri2}), assuming that
the magnetic unit cells (MUC) are either {\bf (i)}
original square-lattice unit cell or
{\bf (ii)} $2 \times 2$ of the original unit cell.
The dimension of the (real-valued) parameter space
in each case becomes
{\bf (i)} $32(+3)$ and {\bf (ii)} $128(+12)$. Starting
from a randomly chosen initial point in these multiple
dimensional parameter spaces, we perform the Newton-Raphson
method, only to reach a certain local minimum
of the mean-field free energy $E^{\rm mf}$
(per the magnetic unit cell);
\begin{align}
E^{\rm mf} \equiv &
\frac{J_1}{4} \sum_{\langle jl\rangle\in {\rm MUC}}
\big( |{\bm E}_{jl}|^2 + |{\bm D}_{jl}|^2 \big) \nn \\
&+ \frac{J_2}{4} \sum_{\langle\langle jl \rangle\rangle \in {\rm MUC}}
\big( |\chi_{jl}|^2 + |\eta_{jl}|^2 \big)
 \nn \\
& - \frac{1}{16\pi^2}\sum_{\alpha=1}^{\nu}
\int \int_{\rm MBZ} dk_x dk_y\ |\lambda_{\alpha}|, \label{mfe}
\end{align}
with {\bf (i)} $\nu=4$ or {\bf (ii)} $\nu=16$.
Here, the summation
over $jl$ is taken within each
magnetic unit cell and $\lambda_{\alpha}$
denotes the spinon energy band.
We have repeated
this procedure from 50 times to 300 times for
each parameter point, i.e.
$(J_1,J_2) = (\sin\theta,\cos\theta)$ with $0\le\theta\le \frac{\pi}{2}$.
In this way, we enumerated various spin-triplet RVB ansatzes.

Throughout this extensive search, we found basically
three distinct RVB ansatzes having both spin-triplet
link-variable on each NN bond and spin-singlet
link-variable on each NNN bond. All of these three
do not break any translational symmetries of the
original unit cell, i.e. $T_x$
and $T_y$.

\subsubsection{$Z_2$ Balian-Werthamer state}
\begin{figure}[tbh]
    \includegraphics[width=80mm]{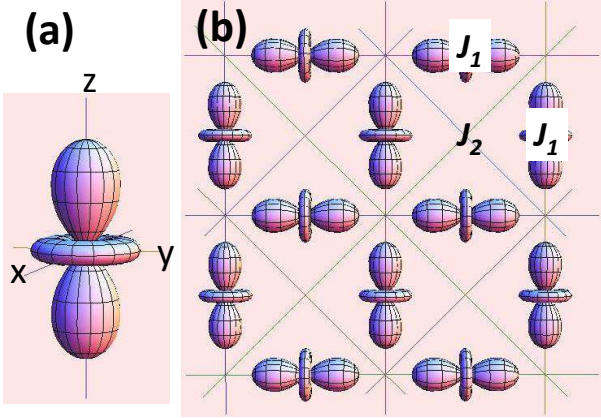}
\caption{(color online) (a) $2z^2 -x^2-y^2$ type quadrupole moment formed by two $S=1/2$
spins on each bond.
(b) $J_1$-$J_2$ model and the configuration of the quadrupole moments
on bonds in
the $Z_2$ BW state [see Eqs.\ (\ref{eq:quadrupole1}) and
(\ref{eq:quadrupole2})].}
\label{fig:quadrupole}
\end{figure}

The first one is 
a sort of the Balian-Werthamer (BW) state\cite{BW} where the 
${\bm d}$-vector
on the NN $x$-link is perpendicular to that on the $y$-link,
\begin{align}
&
U^{\rm tri}_{\langle j,j+\hat{x}\rangle,\mu}
= i\delta_{\mu 1}D \sigma_2, \ \ \
U^{\rm tri}_{\langle j,j+\hat{y}\rangle,\mu}
= i\delta_{\mu 2} D \sigma_2, \nn \\
&
U^{\rm sin}_{\langle j,j+\hat{x}\pm \hat{y}\rangle}
= \chi \sigma_3 \pm \eta \sigma_1, \ \ \
ia_{\nu} = 0.
\label{z2bw}
\end{align}
`$D$', `$\chi$' and `$\eta$' above
correspond to the real parts of
Eqs.~(\ref{tri3-1}) and (\ref{sin3-12}), respectively.
This RVB state exhibits  the same antiferro-type configuration  
of quadrupolar moments as the bond nematic state 
found in Ref.\ \onlinecite{sms}. Namely, the nematic
order parameters on NN bonds show
\begin{equation}\label{eq:quadrupole1}
Q_{jl,11}= - \frac{2}{3}D^2,\ \ \ Q_{jl,22}=Q_{jl,33}=\frac{1}{3} D^2
\end{equation}
for the $x$-direction and 
\begin{equation}\label{eq:quadrupole2}
Q_{jl,22}=-\frac{2}{3}D^2,\ \ \ Q_{jl,11}=Q_{jl,33}=\frac{1}{3}D^2 
\end{equation}
for the $y$-direction, where $Q_{jl,\mu\nu}=0$ for 
$\mu\ne\nu$ (see Fig.~\ref{fig:quadrupole}).
While this mean-field ansatz breaks the mirror symmetry $P_{xy}$
which interchanges $x$-link and $y$-link, it
is invariant under the following combined
symmetry and gauge transformations: $G_x T_x$,
$G_y T_y$, $G_{P_x} P_x$, $G_{P_y} P_y$,
$G_{P^{\prime}_{xy}} P^{\prime}_{xy}$ and
$G_{{\cal T}} {\cal T}$. The respective gauge
transformations read
\begin{align}
&
G_{x} =  G_y = \sigma_0, \ \ \ G_{P_x} = i\sigma_1 (-1)^{j_x},
\ \ \
G_{P_y} = i\sigma_1 (-1)^{j_y}, \nn \\
&
G_{P^{\prime}_{xy}} = i\sigma_2(-1)^{j_y}, \ \ \
G_{\cal T} = (-1)^{i_x + i_y}.  \label{psg?z2bw}
\end{align}
Here ${\cal T}$ 
refers to the time-reversal symmetry,
while $P^{\prime}_{xy}$ stands for the mirror symmetry $P_{xy}$
accompanied by an appropriate spin-rotation about the $3$-axis
by $\pi/2$.

Provided $\eta \chi \ne 0$, the
ansatz supports two non-collinear $SU(2)$ gauge fluxes,
\begin{align}
U^{\rm tri}_{\langle j,j+\hat{x}\rangle,1}
U^{\rm tri}_{\langle j+\hat{x},j+\hat{x}+\hat{y}\rangle,2}
U^{\rm sin}_{\langle j+\hat{x}+\hat{y},j\rangle} &\propto
\chi \sigma_3 + \eta \sigma_1, \label{su2flux1} \\
U^{\rm tri}_{\langle j,j+\hat{x}\rangle,1}
U^{\rm tri}_{\langle j+\hat{x},j+\hat{x}-\hat{y}\rangle,2}
U^{\rm sin}_{\langle j+\hat{x}-\hat{y},j\rangle} &\propto
\chi \sigma_3 - \eta \sigma_1. \label{su2flux2}
\end{align}
Hence it 
is protected from any small gauge fluctuation by finite
Higgs mass. We call this ansatz as the $Z_2$ BW
state. The spinon's band dispersion $\lambda_{\alpha}$ 
of this $Z_2$ state is comprised of two doubly degenerate
bands, both of which are always separated by a finite energy
gap in the entire Brillouin zone, $[-\pi,\pi]\times [-\pi,\pi]$;
\begin{eqnarray}
\lambda_{1,2} \equiv -\lambda_{3,4}
\equiv \big\{A^2(s^2_x + s^2_y) + B^2 c^2_x c^2_y + C^2 s^2_xs^2_y
\big\}^{\frac{1}{2}} \label{z2bwband}
\end{eqnarray}
with $(s_{\mu},c_{\mu}) \equiv (\sin k_{\mu},\cos k_{\mu})$
and
$(2A,B,C)\equiv (J_1 D,J_2\chi,J_2\eta)$.

\subsubsection{$SU(2)$ chiral $p$-wave state}\label{sec:ABM}
The second ansatz we found is the chiral $p$-wave
[Anderson-Brinkman-Morel (ABM)] state,\cite{ABM} in which
all the ${\bm d}$-vectors
on the NN-bonds are collinear, while
the ${\bm d}$-vector on the $x$-link acquires extra
phase $i$ in relative to that on the $y$-link,
\begin{align}
&
U^{\rm tri}_{\langle j,j+\hat{x}\rangle,\mu}
= i\delta_{\mu 3}D \sigma_2, \ \ \
U^{\rm tri}_{\langle j,j+\hat{y}\rangle,\mu}
= i\delta_{\mu 3} D \sigma_1, \nn \\
&
U^{\rm sin}_{\langle j,j+\hat{x}\pm \hat{y}\rangle}
= \chi \sigma_3, \ \ \ ia_{\nu} = 0. \label{su2abm}
\end{align}
Namely, two `$D$' appearing in the
first line stand for the real and imaginary part of
the ${\bm d}$-vector respectively. Because of this
relative phase factor, this ansatz has its
fermionic band-dispersion fully gapped in the
whole momentum space;
\begin{eqnarray}
\lambda_{1,2} = - \lambda_{3,4} = \lambda_k =
\big\{A^2(s^2_x+s^2_y)
+ B^2c^2_x c^2_y \big\}^{\frac{1}{2}}. \label{su2abmband}
\end{eqnarray}
In this state, all
NN bonds have the same ferro-nematic order
$Q_{jl,33}=-\frac{2}{3}D^2$, $Q_{jl,11}=Q_{jl,22}=\frac{1}{3}D^2$.

The IGG of this chiral $p$-wave state contains
the following three continuous gauge symmetries:
\begin{align}
\{ e^{i(-1)^{j_x+j_y}\theta\sigma_3},
e^{i(-1)^{j_x}\theta\sigma_1},
e^{i(-1)^{j_y}\theta\sigma_2}|\theta\in[0,2\pi)\}.
\label{rot1}
\end{align}
Correspondingly, the low-energy effective theory in
the gauge (non-magnetic) part consists of three
maxwell forms around $q=(\pi,\pi)$, $(\pi,0)$ and $(0,\pi)$ 
respectively. Namely, above continuous gauge symmetries 
require that the following three types of non-magnetic
deformations constitute the $U(1)$ gauge
invariant effective actions:
\begin{align}
\big\{ U^{\rm tri}_{jl,\mu},U^{\rm sin}_{jl}
\big\} &=
\big\{\bar{U}^{\rm tri}_{jl,\mu},
\bar{U}^{\rm sin}_{jl}\big\} e^{i(j-l)_{\alpha}(-1)^{l_x+l_y}
a_{\alpha}(l,\tau)\sigma_{3}}
,\nn\\
a^{3}_{j,\tau} &= (-1)^{j_x+j_y} a_{0}(j,\tau), \label{3} \\
\big\{ U^{\rm tri}_{jl},U^{\rm sin}_{jl}
\big\} &=
\big\{\bar{U}^{\rm tri}_{jl,\mu},
\bar{U}^{\rm sin}_{jl}\big\} e^{i(j-l)_{\alpha}(-1)^{l_x}
a_{\alpha}(l,\tau)\sigma_{1}}
,\nn\\
a^{1}_{j,\tau} &= (-1)^{j_x} a_{0}(j,\tau), \label{1} \\
\big\{ U^{\rm tri}_{jl,\mu},U^{\rm sin}_{jl}
\big\} &=
\big\{\bar{U}^{\rm tri}_{jl,\mu},
\bar{U}^{\rm sin}_{jl}\big\} e^{i(j-l)_{\alpha}(-1)^{l_y}
a_{\alpha}(l,\tau)\sigma_{2}}
,\nn\\
a^{2}_{j,\tau} &= (-1)^{j_y} a_{0}(j,\tau). \label{2}
\end{align}
Though these three types of gauge fluctuations
are not suppressed by finite Higgs mass, the ansatz itself
is still protected by the
so-called Chern-Simon mechanism.\cite{CS,Pisarski,wen3,khPW,volovik-yakovenko,wen,fradkin}

To see this, notice that the ansatz 
(\ref{su2abm})
breaks  all the mirror symmetries
$P_x$, $P_y$, $P_{xy}$ and the time-reversal symmetry
${\cal T}$. Instead, it is invariant
only under these mirror symmetries
accompanied by the time-reversal
symmetry $G_{P{\cal T}}\cdot P{\cal T}$
or under the spatial inversion
symmetry $G_{R_{\pi}}R_{\pi}$. 
The respective gauge transformations are given by
\begin{align}
G_{P_{y}{\cal T}}
&= \sigma_0, \ \
G_{\cal R_{\pi}} = G_{P_{x}{\cal T}} = (-1)^{j_x+j_y}, \nn \\ 
G_{P_{xy}{\cal T}} &= i(-\sigma_3)^{j_x+j_y}. 
\label{psg?abm}
\end{align}
This magnetic point group clearly allows the spontaneous
Hall conductance of the `spinon',
like in the chiral spin state.\cite{wen3,khPW,volovik-yakovenko}
In fact, corresponding to the three continuous
gauge symmetries given in Eq.~(\ref{rot1}),
we have three conserved `charges', all of which are
accompanied by finite quantized transverse conductance
$\sigma_{xy} = \frac{2}{2\pi}$.
As a result, the effective actions around $q=(\pi,\pi)$, $(0,\pi)$
and $(\pi,0)$ acquire the Chern-Simon term in addition to
the maxwell form,\cite{volovik-yakovenko,wen3,khPW,wen,fradkin}
\begin{eqnarray}
F_{\rm gauge} \equiv \int dx^2 d\tau
\frac{\sigma_{xy}}{2} a_{\mu}\partial_{\nu}a_{\lambda}
\epsilon_{\mu\nu\lambda}\ +  ({\rm maxwell}\ {\rm form}). \nn
\end{eqnarray}
This Chern-Simon term endows the apparently massless
gauge boson with a finite energy gap.\cite{Pisarski}

\subsubsection{$Z_2$ collinear state}
The third stable ansatz we found is
the `collinear' state, where all
${\bm d}$-vectors are pointing to the same
direction,
\begin{eqnarray}
U^{\rm tri}_{\langle j,j+\hat{x}\rangle,\mu}
= U^{\rm tri}_{\langle j,j+\hat{y}\rangle,\mu}
= i \delta_{\mu 3}D\sigma_2, \nn \\
U^{\rm sin}_{\langle j,j+\hat{x}\pm \hat{y}\rangle}
= \chi \sigma_3 \pm \eta \sigma_1,
\ \ \ ia^{1}_{j,\tau} \ne 0, \label{z2collin}
\end{eqnarray}
showing ferro-nematic order $Q_{jl,33}=-\frac{2}{3}D^2$,
$Q_{jl,11}=Q_{jl,22}=\frac{1}{3}D^2$.
Although having the same spin-quadrupolar moment
as the previous one, this collinear
ansatz is a distinct quantum order state from the
$SU(2)$ chiral $p$-wave state.
It preserves mirror
symmetries as well as the time-reversal symmetry.
In fact, one can see that all the discrete
symmetries of the original square lattice are
recovered, when combined with the following gauge
transformations:
\begin{align}
&
G_{x} = G_y = \sigma_0, \ \ \
G_{P_x} = i\sigma_1(-1)^{j_x}, \ \ \
G_{P_y} = i\sigma_1(-1)^{j_y}, \nn \\
&
G_{P_{xy}} = 1, \ \ \
G_{\cal T} = (-1)^{j_x+j_y}. \label{psg?z2cl}
\end{align}
Having the non-collinear $SU(2)$ gauge fluxes as in Eqs.~(\ref{su2flux1})
and (\ref{su2flux2}), all the gauge fluctuations around this ansatz are
suppressed by finite Higgs mass.
We hence call this state as $Z_2$ collinear state.

\subsection{Phase diagram}
\begin{figure}[ht]
\begin{center}
\includegraphics[width=0.46\textwidth]{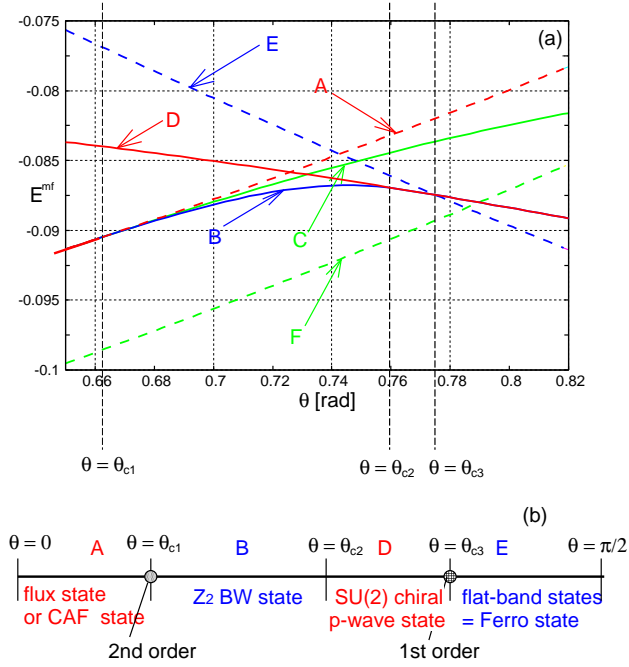}
\end{center}
\caption{(color online) (a) Mean-field energies (per site) of the various
ansatzes in the $S=1/2$ square lattice ferromagnetic $J_1$-$J_2$
model. Note that
$(J_1,J_2)\equiv |J|(\sin\theta,\cos\theta)$, where the
energy unit is taken to be $|J|$. The Blue line (labeled as $B$)
is for the Balian-Werthamer (BW) state, which is the $Z_2$ state for
$\theta_{c1} < \theta < \theta_{c2}$ and which reduces
to the $U(1)$ state for $\theta_{c2} < \theta $. The green
line (labeled as $C$) is for the $Z_2$ collinear state, while the
red line (labeled as $D$) stands for the $SU(2)$ chiral $p$-wave state.
The red dotted line (labeled as $A$)
is the doubled $\pi$-flux state, where both the $A$-sublattice
and the $B$-sublattice support $\pi$-flux states respectively
[see Eq.~(\ref{flux})].
The Blue dotted line (labeled as $E$) is
for a set of `flat-band' states
($E^{\rm mf}_{\rm flat}= -\frac{1}{8}|J|\sin\theta$), all
of which give the same best mean-field energy  for
$\theta_{c3}<\theta$. The green dotted line (labeled as $F$)
is for the staggered dimer state
($E^{\rm mf}_{\rm dimer}= -\frac{1}{8}|J|\cos\theta$),
where both the $A$-sublattice and the $B$-sublattice
support staggered dimer states respectively [see
Eq.~(\ref{dimer}) for its example].
These isolated dimer states are known to be overcome
energetically by the doubled $\pi$-flux state,\cite{Gros-FCZhang} 
when they are projected onto the physical Hilbert space.
Since the $Z_2$ BW state is composed on
the top of the $\pi$-flux state, this staggered dimer state
is also expected to be overcome by the {\it projected} $Z_2$ BW
state. 
(b) Expected mean-field phase diagram in the
intermediate coupling region. The transition at $\theta_{c1}$
is the 2nd order, since the magnetic space group
of the $Z_2$ BW state belongs to that of the $\pi$-flux state.
On the other hand, the transition at $\theta_{c3}$ is
the 1st order at the mean-field level, which one can see
directly from the Figure~(a). }
\label{mfsum}
\end{figure}

The mean-field energy for these three ansatzes are
plotted in Fig.~\ref{mfsum}(a)
with $(J_1,J_2) \equiv J(\sin\theta,\cos\theta)$.
Let us begin with the lowest energy mean-field solution in the
well-studied limit, $J_2 \gg J_1$.
In the strong $J_2$ limit, our model reduces to
the {\it two decoupled} antiferromagnetic square
lattice, so that the knowledges of the saddle-point solutions
in this limit have been well-established.\cite{pwa,kma,Dombre,Gros-FCZhang,
liang,read-sachdev2,wen,fradkin,review} Namely, the
$\pi$-flux state defined on each square lattice,
\begin{eqnarray}
U^{\rm sin}_{\langle j,j+\hat{x}\pm \hat{y} \rangle} =
\chi \sigma_3 \pm \eta \sigma_1, \ \ \
U^{\rm tri}_{\langle j,j+\hat{\nu}\rangle,\mu}
= ia^{\mu}_{j,\tau} = 0 \label{flux}
\end{eqnarray}
with $\chi = \eta$, becomes
global minimum, when the magnetic unit cell (MUC) is
restricted to the original square lattice unit cell.
On the other hand, when the MUC is enlarged
up to the 2 $\times$ 2, the global minimum state becomes
the staggered dimer state introduced on each decoupled
square lattice, e.g.
\begin{align}
& U^{\rm sin}_{\langle j,j+\hat{x} + \hat{y} \rangle}
= U^{\rm sin}_{\langle j+\hat{x}, j+ \hat{y} \rangle} = \chi \sigma_3, 
\ \ U^{\rm tri}_{\langle j,j+\hat{\nu}\rangle ,\mu}
= ia^{\mu}_{j,\tau} = 0 \nn \\ 
& U^{\rm sin}_{\langle j+\hat{x},j+2\hat{x} + \hat{y} \rangle}
= U^{\rm sin}_{\langle j+2\hat{x},j+\hat{x} + \hat{y} \rangle}
= 0, \nn \\ 
& U^{\rm sin}_{\langle j+\hat{y}, 
j+\hat{x} + 2\hat{y} \rangle} = U^{\rm sin}_{\langle j+\hat{x} + \hat{y},
j+2\hat{y} \rangle} =  0, \nn \\ 
& U^{\rm sin}_{\langle j+\hat{x}+\hat{y},
j+2\hat{x}+2\hat{y} \rangle} = U^{\rm sin}_{\langle j+2\hat{x} + \hat{y},
j+\hat{x}+ 2\hat{y} \rangle} = 0. \label{dimer}
\end{align} 
However, using the variational Monte Carlo (VMC) calculations,
Gros and his co-workers\cite{Gros-FCZhang}
have demonstrated that,
when projected onto the original
(spin) Hilbert space, the $\pi$-flux state eventually
wins over this isolated dimer state.
In fact, it is well-established\cite{review} that the projected
$\pi$-flux state gives the second best variational energy in the
strong $J_2$ limit (the best variational estimate
is obtained from the Neel order state\cite{liang}).

When increasing the NN ferromagnetic interaction $J_1$,
a finite spin-triplet ansatz continuously develops on the
top of this $\pi$-flux state, while simultaneously
the parameters $\eta$ start to deviate from $\chi$, i.e.,
$\eta \ne \chi$. This leads to either $Z_2$ BW state or
$Z_2$ collinear state for
$\theta_{c1} \equiv 0.66<\theta$.
Thus, the transitions from the $\pi$-flux state to
these two $Z_2$ states are both the second order at the 
mean-field level. Energetically speaking, the 
$Z_2$ BW state gives a slightly 
lower mean-field energy than that of the 
$Z_2$ collinear state. 

Notice also that 
these two $Z_2$ states 
are clearly pre-emptted by the staggered dimer state, 
Eq.~(\ref{dimer}),  
at the mean-field level [see Fig.~\ref{mfsum}(a)].   
Observing the situation in the strong $J_2$ limit, 
however, one can naturally expect that, when projected 
onto the physical (spin) Hilbert space, both $Z_2$ 
states would win over this isolated dimer state in the 
case of a finite $J_1$. Namely, since our $Z_2$ states are 
constructed based on the decoupled $\pi$-flux states [compare 
Eqs.~(\ref{z2bw},\ref{z2collin}) with Eq.~(\ref{flux})],  
they would certainly acquire substantial resonance energies 
in the same way as the $\pi$-flux state does. 
On the other hand, being factorisable, any isolated dimer state 
cannot gain such resonance energies, irrespective of finite  
ferromagnetic exchange interactions. Moreover, 
Fig.~\ref{mfsum}(a) indicates that the $Z_2$ BW asatz is quite 
energetically tunable in the presence of the ferromagnetic 
exchange interaction.  
Thus, we presume that the  
$Z_2$ BW state finally dominates in this 
intermediate coupling region, 
$\theta_{c1}\equiv 0.66<\theta$.   

When $\theta_{c2}\equiv 0.76<\theta$,
this $Z_2$ BW state reduces to the $U(1)$ state having
no finite $\eta$. Namely, with $\eta=0$, two $SU(2)$ gauge
fluxes given in Eqs.~(\ref{su2flux1}) and (\ref{su2flux2})
become collinear with each other. Simultaneously, this $U(1)$
BW state becomes energetically degenerate with
the $SU(2)$ chiral $p$-wave state. Namely, both of them have
precisely the same mean-field band dispersions $\pm \lambda_k$
[compare Eq.~(\ref{su2abmband}) with Eq.~(\ref{z2bwband}) having $\eta=0$].

This $U(1)$ BW state is destroyed by the
infinitesimally small gauge fluctuation.
Namely, in the absence of finite $\eta$,
the non-magnetic deformations defined in Eq.~(\ref{t1})
constitute 
the following maxwell form around $q=(\pi,\pi)$,
\begin{eqnarray}
F_{\rm gauge} = \int^{\beta}_{0} d\tau \int d^2x 
 \{u {\bm e}^2 + \frac{1}{2} K b^2\}
+ \cdots \nn
\end{eqnarray}
where $e_{\alpha}$ ($\alpha=1,2$) and 
$b$ are defined, from Eqs.~(\ref{t1},\ref{t1-1}),  
as $e_{\alpha}(j,\tau) \equiv (-1)^{j_x+j_y}
(\partial_{\tau} a_{\alpha} - \partial_{\alpha}
a_{0})$ and $b(j,\tau) \equiv (-1)^{j_x+j_y}
(\partial_{2} a_{1} - \partial_{1}
a_{2})$. Since the fermionic excitations
are fully gapped even without $\eta$ [see Eq.~(\ref{z2bwband})],
this maxwell form is free from any
dissipation effect,\cite{il} e.g.
\begin{eqnarray}
u = \int\int_{[-\pi,\pi]^2}
d^2k  \frac{A^4s^2_yc^2_x + A^2B^2c^2_y(1+s^2_xs^2_y)}
{16\pi^2 \lambda^5_k} > 0. \nn 
\end{eqnarray}
Having the time-reversal
symmetry [see Eq.~(\ref{psg?z2bw})], the massless gauge
fluctuation is not suppressed by the Chern-Simon term
either.\cite{note} Consequently, infinitesimally small 
fluctuations of this type of gauge fields 
lead the $U(1)$ BW state into 
a confining phase having no gapped free spinon in 
its excitation. More specifically, those space-time 
instantons (monopoles)
which are allowed by the corresponding compact QED action, 
$\int d\tau \int d^2x  
\{u{\bm e}^2 - K \cos 
(\epsilon_{\alpha\beta}\Delta_{\alpha} a_{\beta})\}$,  
proliferate in the $2+1$ dimensional space, \cite{polyakov}
lowering a certain magnetic symmetries enumerated 
in Eq.~(\ref{psg?z2bw}).\cite{read-sachdev1} 
To capture the resulting magnetic space group of the 
confining phase, one generally need to identify the 
quantum number carried by this monopole creation 
field. \cite{ran-v-l,alicea}   

For $\theta_{c3} \equiv 0.775<\theta$,
these two degenerate ansatzes, -- $SU(2)$ chiral $p$-wave state 
and $U(1)$ BW state --, are further overcome
(energetically) by another ansatz, which we dubbed as the
`flat-band' states,
\begin{eqnarray}
U^{\rm sin}_{jl} = 0, \ \
U^{\rm tri}_{jl} \ne 0, \ \ ia^{\nu}_{j,\tau} \ne 0.  
\label{flat}
\end{eqnarray}
These `flat-band' states do
not have any finite
singlet ansatzes anymore and keep on giving the lowest
mean-field energy ($E^{\rm mf}_{\rm flat}= -\frac{1}{8}J_1$)
for the remaining ferromagnetic side,
$\theta_{c3}<\theta<\frac{\pi}{2}$.
However, these `flat-band' states do not necessarily refer to
a specific configuration of the spin-triplet ansatzes.
Instead, they refer to a {\it group} of the states all of
which give precisely the same mean-field energy.
For example, these `flat-band' states include the following parameterization
of the spin-triplet ansatz:
\begin{align}
&\left[
\begin{array}{ccc}
{\bm E}^{\prime\prime}_{\langle j,j+\hat{x}\rangle}&
{\bm D}^{\prime\prime}_{\langle j,j+\hat{x}\rangle}&
{\bm D}^{\prime}_{\langle j,j+\hat{x}\rangle} 
\end{array}
\right] = \alpha \ {\bm n}_1 \cdot {\bm m}^{T}_1,  \nn \\
&\left[ 
\begin{array}{ccc}
{\bm E}^{\prime\prime}_{\langle j,j+\hat{y}\rangle}&
{\bm D}^{\prime\prime}_{\langle j,j+\hat{y}\rangle}&
{\bm D}^{\prime}_{\langle j,j+\hat{y}\rangle} 
\end{array}
\right] = \beta \ {\bm n}_1 \cdot {\bm m}^{T}_2,  \nn \\
&{\bm E}^{\prime}_{\langle j,j+\hat{x}\rangle} = \alpha \ {\bm n}_2,  
\ \ {\bm E}^{\prime}_{\langle j,j+\hat{y}\rangle} = 
\beta \ {\bm n}_3, \nn \\ 
&{\bm n}^{T}_1 \cdot {\bm n}_2 = {\bm n}^{T}_2 \cdot {\bm n}_3 
= {\bm n}^{T}_3 \cdot {\bm n}_1 = {\bm m}^{T}_1 \cdot
{\bm m}_2 = 0, \label{gsm1}
\end{align}
where $\alpha^2+\beta^2=\frac{1}{4}$, and ${\bm n}_j$ 
and ${\bm m}_j$ can be arbitrary unit
vectors that observe Eq.~(\ref{gsm1}).
Here $E^{\prime}_{jl,\mu}$ and $D^{\prime}_{jl,\mu}$
stand for the real part of $E_{jl,\mu}$ and $D_{jl,\mu}$,
respectively, while $E^{\prime\prime}_{jl,\mu}$ and
$D^{\prime\prime}_{jl,\mu}$
are their respective imaginary parts.
Thus, only the first one is parity even
$E^{\prime}_{jl,\mu} = E^{\prime}_{lj,\mu}$, while
the others are odd,
$E^{\prime\prime}_{jl,\mu}
= - E^{\prime\prime}_{lj,\mu}$
and $D_{jl,\mu} = - D_{lj,\mu}$.
Bearing these in mind, one can easily see that  
this mean-field ansatz always gives the two doubly 
degenerate spinon bands, which are totally {\it flat} 
in the entire Brillouin zone, 
\begin{eqnarray}
\lambda_{1,2} = - \lambda_{3,4}
= \frac{J_1}{4}. \nn
\end{eqnarray}
Because of this feature, all
the spin-triplet ansatzes parameterized by
Eq.~(\ref{gsm1}) give the same mean-field energy
(per site) $E^{\rm mf}_{\rm flat} =
- \frac{1}{8}J_1$. 

The emergence of these `huge' numbers of
`flat-band' states in the strong $J_1$ limit
reflects the fact that
the ground-state order parameter of any
Heisenberg ferromagnet (total spin moment)
and the corresponding spin Hamiltonian
are {\it simultaneously  diagonalizable}.
When projected onto the physical (spin)
Hilbert space, we expect that these flat-band 
states reduce to a fully 
polarized state (ferromagnetic state).

Observing Fig.~\ref{mfsum},
please notice 
that our $Z_2$ BW phase appears in
larger $J_2$ region in comparison with
the previous ED studies.
Namely, Fig.~\ref{mfsum} indicates that its 
phase boundaries are given by
$J_1:J_2=1:1.29$ at $\theta = \theta_{c1}$ and
$J_1:J_2=1:1.05$ at $\theta = \theta_{c2}$, while
$d$-wave bond nematic order phase was
found in $0.4\lesssim J_2/J_1\lesssim 0.6$ in the previous
finite-size studies.\cite{sms}
This discrepancy simply stems from the so-called
`factor $3$' difference, often encountered between
the Hartree-Fock (HF) spin-singlet ansatz and the HF spin-triplet
ansatz. If one employed a more numerics-oriented
formulation,\cite{UL} $\frac{J_{\rm 1}}{4}$ appearing
in Eq.~(\ref{tri2}) is
replaced by $\frac{J_{\rm 1}}{8}$, while
$\frac{J_{\rm 2}}{4}$
is replaced by $\frac{3J_{\rm 2}}{8}$. Consequently,
we have $J_1:J_2=1:0.43$ ($\theta = \theta_{c1}$) and
$J_1:J_2=1:0.36$ ($\theta = \theta_{c2}$),
which would be relatively comparable with the previous ED
result. More quantitative comparison, however, requires
the variational Monte Carlo studies based on
these spin-triplet ansatzes.

In summary, we have argued that three spin-triplet RVB
ansatzes --- $Z_2$ and $U(1)$ BW states and $SU(2)$
chiral $p$-wave state ---
become the lowest mean-field states
in the intermediate coupling region,
$J_1 \simeq 2J_2$  [see Fig.~\ref{mfsum}(b)].
Among them, both the $Z_2$ BW state and the $SU(2)$ chiral
$p$-wave state are stable against any (infinitesimally)
small gauge fluctuation, while in the $U(1)$
BW state the effect of gauge fluctuation
is crucial, making spinons confined.
Using Eq.~(\ref{dir}), one can easily see
that the BW states show the $d$-wave
bond-type spin quadrupolar order precisely
as in Eq.~(\ref{dwave}).


\subsection{Magnetic excitations in the BW states}
Here we briefly discuss magnetic excitations
in the $Z_2$ BW state.
The `low-energy' excitation around the
$Z_2$ BW state is composed of three parts; {\bf (i)}
gapped non-magnetic excitations (gauge bosons),
{\bf (ii)} gapless magnetic
excitations (Goldstone bosons), and
{\bf (iii)} gapped fermionic ($\Psi$-field)
individual excitations.
The gapped gauge boson plays only a subdominant role
in the spin-structure factor, while the latter two
contribute significantly to magnetic excitations.
Up to the Hartree-Fock
level, one can easily see that the gapped fermionic
excitation constitutes the continuum spectrum above
$\omega>\max(J_1|D|,J_2|\chi|)$. When one further takes 
into account the random-phase approximation terms,\cite{moriya,further} 
the gapless bosonic dispersions emerge
below this spinon continuum, whose low-energy limit
can be described by 
the matrix-formed non-linear $\sigma$ model,
\begin{eqnarray}
F_{\rm magnetic} = \sum_{\mu=\tau,x,y}
{\rm Tr}\big[\hat{\Lambda}_{\mu}\partial_{\mu}
\hat{R}^{-1}\partial_{\mu}\hat
{R}\big]. \label{nlsm}
\end{eqnarray}
Namely, the $3 \times 3$ matrix $\hat{R}$ is nothing but
the spatio-temporally varying rotational matrix of
the director vector used in Eq.~(\ref{rotat}).
The symmetry argument\cite{delamotte} dictates that
the diagonal matrices $\hat{\Lambda}_{\mu}$
generally take the following form:
\begin{eqnarray}
&&\hspace{-0.7cm}
\big\{
\hat{\Lambda}_{\tau},
\hat{\Lambda}_{x},\hat{\Lambda}_y \big\}
\equiv \nn \\
&&\hspace{-0.4cm}\Bigg\{
\left[\begin{array}{ccc}
c_{0} & & \\
& c_{2} & \\
& & c_2 \\
\end{array}\right],
\left[\begin{array}{ccc}
c_1 & & \\
& c_3 & \\
& & c_4 \\
\end{array}\right],
\left[\begin{array}{ccc}
c_1 & & \\
& c_4 & \\
& & c_3 \\
\end{array}\right] \Bigg\}, 
\end{eqnarray}
where the director coplanar plane was taken 
to be the $2$-$3$ plane. 
In terms 
of the semiclassical (gradient) expansion, one 
can directly calculate their respective 
coupling constants:
\begin{align}
c_0 &\equiv 0, \ \ c_1 \equiv \int\int_{[-\pi,\pi]^2}d^2k\ 
\frac{A^4s^2_yc^2_x}{64\pi^2 \lambda^3_k}, \nn \\  
c_2  &\equiv \int 
\int_{[-\pi,\pi]^2}d^2k \ \frac{A^2(s^2_x + s^2_y)}{64\pi^2 \lambda^3_k}, \nn \\
c_3 & \equiv  \int 
\int_{[-\pi,\pi]^2}d^2k \  \bigg\{
\frac{\mathcal{J}_1A^2s^2_x}{8\pi^2} - 
\frac{A^4s^2_yc^2_x}{64\pi^2 \lambda^3_k} \bigg\}, \nn \\ 
c_4 & \equiv  \int 
\int_{[-\pi,\pi]^2}d^2k \ \bigg\{
\frac{\mathcal{J}_1A^2s^2_y}{8\pi^2} - 
\frac{A^4s^2_yc^2_x}{64\pi^2 \lambda^3_k} \bigg\}, \nn 
\end{align}
with $\mathcal{J}_1 \equiv 4^{-1}\lambda^{-1}_k
\{(\partial_{k_x} {\bm n}^{T})
(\partial_{k_x} {\bm n})
+ 2^{-1}\lambda^{-1}_k 
\partial^2_{k_x} \lambda_k\}$, and 
${\bm n} \equiv 2^{-1}\lambda^{-1}_k
(2As_x,2As_y,Bc_xc_y,Cs_xs_y )$.  
In addition to these massless
excitations, we could also have several gapped (`optical')
magnetic modes, provided that they are not damped by the
spinon individual excitations.\cite{moriya,further}
One might also expect a certain characteristic behavior of
the spectral weight themselves. In fact, Tsunetsugu
{\it et al.}\cite{tsune} and Lauchli {\it et al.}\cite{lmp} 
 demonstrated that the spin-structure
factor in the site-nematic ordered state exhibits the
{\it vanishing} spectral intensities of the Goldstone
modes around the $\Gamma$-point.


\section{summary and open issues}

In this paper, we have introduced the spin-triplet
slave-boson formulation as a mean-field theory for
the bond-type spin nematic state, which
was described as the spin-triplet RVB state.  Namely,
the ${\bm d}$-vectors of the spin-triplet RVB ansatz
constitute the quadrupolar
order, while the combination of
the spin-triplet and singlet link
variables on the same link leads to the
vector chiral order.

When applied to the $S=\frac{1}{2}$
square-lattice frustrated ferromagnetic Heisenberg
model, our spin-triplet slave-boson analysis gives
two non-trivial stable spin-triplet RVB ansatzes in the
intermediate coupling region around $J_1:J_2 \simeq 1:0.4$.
One is the $Z_2$ BW state stabilized by
the Anderson-Higgs mechanism, while the other is
the $SU(2)$ chiral $p$-wave state protected by the
Chern-Simon mechanism. Our slave-boson analysis also found
an unstable $U(1)$ BW state as a mean-field solution,
which possibly gives a route to the realization of spinon
confined quadrupolar ordered states with a certain 
symmetry reduction.  
The projective symmetry
group of the $Z_2$ BW state 
as well as the $U(1)$ BW state is consistent with the magnetic
space group of the $d$-wave bond-type spin nematic state discussed in
Ref.\ \onlinecite{sms}. Both of them exhibit the 
antiferro-type configuration 
of the bond quadrupolar moment shown 
in Fig.\ \ref{fig:quadrupole}.

Contrary to a naive expectation, our BW state is
classified into a `weak topological (ordinary) insulator' 
instead of the `strong topological insulator'  
defined in the recent literatures.\cite{km1,z2qsh,fkm,QHZ,SRFL} 
Physically speaking, such a `weak topological insulator (WTI)' 
is accompanied either by no spinon edge states at all  
or by even numbers of the helical edge states.
To see that it is indeed a `WTI',
one can first deform this $Z_2$ ansatz
into the $U(1)$ ansatz $(\eta \rightarrow 0)$. 
Since the fermionic dispersion remains gapped, 
the ${\cal Z}_2$ topological index associated
with the filled spinon band\cite{km1,fkm} 
is also unchanged. After reaching the 
simpler $U(1)$ ansatz, let us then
utilize the Fermi surface argument recently
introduced by Sato.\cite{sato}
His argument relates the ${\cal Z}_2$ topological
index in the superconducting state $(D \ne 0)$ with the
Fermi surface topology in the
corresponding `normal' state ($D=0$).
That is, if a Fermi surface in the normal state surrounds
odd/even numbers of the time-reversal invariant
momentum (TRIM) points, the BW state constructed
on top of this normal state is accompanied by
non-trivial/trivial ${\cal Z}_2$ topological index.
Since our normal state is composed of {\it two}
decoupled u-RVB states at $\eta=0$, the resulting Fermi
surface clearly surrounds {\it two} time-reversal
symmetric $k$-points, i.e. $(0,0)$ and $(\pi,\pi)$.
Thus, our $Z_2$ BW state should be classified into the
`WTI' (-- ${\cal Z}_2$ even class --).

In the followings, we will enumerate several open 
issues and possible extensions of the current work. 
The most immediate open issue is to identify the 
magnetic space group of the confining phase 
proximate to the $U(1)$ BW state based on 
the monopole field studies.\cite{alicea,ran-v-l} Namely, such 
an analysis gives several complementary 
informations to the direct ED studies of the  
original spin model.\cite{sms} 

The fate of the $SU(2)$ chiral $p$-wave state observed at
$\theta_{c2} < \theta < \theta_{c3}$ is not so 
clear either, although we have argued its stability 
against any (infinitesimally) small gauge fluctuation. 
Namely, previous exact diagonalization studies of 
the $SU(2)$ spin model did not find any $\cal T$-symmetry 
breaking ferro-nematic states between the 
$d$-wave bond-nematic state and ferromagnetic 
state. In fact, it is also possible that, 
when projected onto the real (spin) Hilbert space,  
the strong gauge fluctuation could wipe out this 
time-reversal breaking ansatz. 

Though we have mainly discussed the quadrupolar 
order in this paper, our formulation can also describe
vector chiral order having no quadrupolar
moment,\cite{ldlst} i.e. $P_{jl,\lambda} \ne 0$
and $Q_{jl,\mu\nu}= 0$. In fact, such vector 
chiral order state was observed in the spin one half 
frustrated Heisenberg model having the ring exchange coupling.\cite{ldlst} 
When applying the current spin-triplet slave-boson 
formulation onto these quantum spin systems, one 
could use the following mean-field parameterization:
\begin{align}
&\left[
\begin{array}{ccc}
{\bm E}^{\prime\prime}_{jl}&
{\bm D}^{\prime\prime}_{jl}&
{\bm D}^{\prime}_{jl} \end{array}
\right] =  \left[\begin{array}{ccc}
{\bm n}_1& {\bm n}_2 &{\bm n}_3 \\
\end{array}\right],  \nn \\
&\left[\begin{array}{ccc}
\chi^{\prime}_{jl} & -\eta^{\prime}_{jl} &
\eta^{\prime\prime}_{jl} \\
\end{array}\right] =
\left[\begin{array}{ccc}
\gamma_1 &
\gamma_2 &
\gamma_3 \\ \end{array}\right], \nn
\end{align}
where $\{{\bm n}_1,{\bm n}_2,{\bm n}_3\}$
are the normalized unit vectors 
orthogonal to one another. Namely, such an ansatz 
gives a finite vector chirality, ${\bm P}_{jl}\equiv
2i\sum_{\alpha=1}^3 \gamma_{\alpha}{\bm n}_{\alpha}$,
without any quadrupolar moments. We generally have 
three alternative ways to parameterize this 
vector chiral order, 
\begin{align}
&\left[
\begin{array}{ccc}
{\bm E}^{\prime}_{jl}&
{\bm D}^{\prime\prime}_{jl}&
{\bm D}^{\prime}_{jl} \\
\end{array}
\right] =  \left[\begin{array}{ccc}
{\bm n}_1& {\bm n}_2 & {\bm n}_3 \\
\end{array}\right],  \nn \\
&\left[\begin{array}{ccc}
\chi^{\prime\prime}_{jl} & -\eta^{\prime}_{jl} &
\eta^{\prime\prime}_{jl} \\
\end{array}\right] =
\left[\begin{array}{ccc}
\gamma_1 &
\gamma_2 &
\gamma_3 \\ \end{array}\right], \nn
\end{align} 
or  
\begin{align}
&\left[
\begin{array}{ccc}
{\bm E}^{\prime\prime}_{jl}&
{\bm E}^{\prime}_{jl}&
{\bm D}^{\prime}_{jl} \\ 
\end{array}
\right] =  \left[\begin{array}{ccc}
{\bm n}_1 & {\bm n}_2 & {\bm n}_3 \\
\end{array}\right],  \nn \\
&\left[\begin{array}{ccc}
\chi^{\prime}_{jl} & \chi^{\prime\prime}_{jl} &
\eta^{\prime\prime}_{jl} \\ 
\end{array}\right] = 
\left[\begin{array}{ccc}
\gamma_1 &
\gamma_2 &
\gamma_3 \\ \end{array}\right], \nn 
\end{align}
or    
\begin{align}
&\left[
\begin{array}{ccc}
{\bm E}^{\prime\prime}_{jl}&
{\bm D}^{\prime\prime}_{jl}&
{\bm E}^{\prime}_{jl} \\
\end{array}
\right] =  \left[\begin{array}{ccc}
\vec{n}_1&\vec{n}_2 &\vec{n}_3 \\
\end{array}\right],  \nn \\
&\left[\begin{array}{ccc}
\chi^{\prime}_{jl} & -\eta^{\prime}_{jl} &
\chi^{\prime\prime}_{jl} \\ 
\end{array}\right] =
\left[\begin{array}{ccc}
\gamma_1 &
\gamma_2 &
\gamma_3 \\ \end{array}\right]. \nn
\end{align}
\begin{acknowledgments}
We acknowledge Takuma Ohashi,
Sung-Sik Lee, Hosho Katsura,
Naoto Nagaosa, Yong Baek Kim,
Leon Balents, Seiji Yunoki, 
Masao Ogata, Nic Shannon, 
Philippe Sindzingre, 
Keisuke Totsuka and Akira Furusaki 
for helpful discussions  
and encouragements. We are especially grateful to
Sung-Sik Lee for clarifying  
the symmetry reduction induced by 
the monopole proliferations, to Keisuke 
Totsuka for clarifying the matrix-formed NL$\sigma$M, 
to Seiji Yunoki for his advice 
on the efficient coding of the Newton-Raphson method.
We are also grateful to Takuma Ohashi for his 
collaboration in Schwinger boson formulation 
in the early stage of this work. 
RS was supported by the
Institute of Physical and Chemical Research (RIKEN) and
TM was supported by Grants-in-Aid for Scientific Research
from the Ministry of Education,
Culture, Sports, Science and Technology (MEXT) of Japan
(Grants No.\ 17071011 and No.\ 20046016).
Part of this work was done during the international 
workshop ``Topological Aspects of
Solid State Physics (TASSP)'', which were  
supported by the Institute for Solid State Physics 
(ISSP), University of Tokyo, Yukawa Institute,  
Kyoto University and I2CAM (with U.S. NSF  
I2CAM International Materials Institute Award, 
Grant DMR-0645461).   
\end{acknowledgments}
\appendix 
\section{large-$N$ frustrated ferromagnetic model}
The mean-field analysis described in this paper 
becomes exact in the large $N$ limit of 
the following action; 
\begin{align}
Z =& \int dU^{\rm sin} dU^{\rm tri}
 d\vec{a}_{\tau}
d\Psi^{a\dagger} d\Psi^{a} \exp
\bigg[- \int^{\beta}_0 d\tau \mathcal{L}\bigg], \label{gauss1} \\
\mathcal{L} =& \frac{1}{2} \sum_{j} {\rm Tr}\big[\Psi^{a\dagger}_j
\big(\partial_{\tau}\sigma_0  + \sum_{\nu=1}^3 ia^{\nu}_{j,\tau}
\sigma_{\nu}\big)\Psi^a_j\big] \nn \\
&\hspace{-.3cm}
-  \frac{J_{1}}{4} \sum_{\langle jl \rangle}
\Big\{N \big(-|{\bm E}_{jl}|^2 - |{\bm D}_{jl}|^2\big) + 
{\rm Tr} 
\big[\Psi^{a\dagger}_j
U^{\rm tri}_{jl,\mu}\Psi^a_{l}\sigma^{T}_{\mu}\big]
\Big\} \nn \\
&\hspace{-.3cm}
-  \frac{J_{2}}{4}  \sum_{
\langle\langle  jl \rangle\rangle}
\Big\{N\big(-|\chi_{jl}|^2 - |\eta_{jl}|^2\big)  + {\rm Tr}
\big[\Psi^{a\dagger}_j
U^{\rm sin}_{jl}\Psi^a_{l}\big]\Big\},  \nn \\
\label{largeN0}
\end{align}
where the summations with respect to the fermion's species 
index $a$ $(=1,\cdots,N)$ were made implicit. The integration  
over the auxiliary fields leads the following large-$N$ 
spin Hamiltonian for frustrated ferromagnets: 
\begin{align} 
{\cal H}_{N} \equiv 
- \frac{J_1}{N} 
\sum_{\langle jl \rangle} \big\{ {\bm S}^{ab}_j 
\cdot {\bm S}^{ba}_{l} +  
\psi^{ab}_{j} \psi^{ba}_{l} \big\} 
+ \frac{J_2}{N} 
\sum_{\langle \langle jl \rangle \rangle} 
{\bm S}^{ab}_j \cdot  
{\bm S}^{ba}_l. \label{largeN}
\end{align}
Note that, in addition to the usual $SP(2N)$ spin 
operators,\cite{note3} we have the density operator which is 
asymmetric in the fermion's species index:    
\begin{align}  
\psi^{ab}  &\equiv \frac{i}{2}
\big( f^{a\dagger}_{\alpha}f^b_{\alpha} 
-f^{b\dagger}_{\alpha} f^a_{\alpha} \big),\ 
{S}^{ab3} \equiv \frac{1}{2} 
\big( f^{a\dagger}_{\uparrow} f^{b}_{\uparrow} 
- f^{b\dagger}_{\downarrow}f^{a}_{\downarrow}\big),  \nn \\
{S}^{ab+}  &\equiv \frac{1}{2} 
\big( f^{a\dagger}_{\uparrow} f^{b}_{\downarrow} 
+ f^{b\dagger}_{\uparrow}f^{a}_{\downarrow}\big), \ 
S^{ab-}\equiv \{S^{ab+}\}^{\dagger}. \label{s+-}
\end{align}
The Hilbert space of this generalized spin Hamiltonian  
is defined as the $SU(2)$-gauge invariant subspace of 
the fermionic Hilbert space.\cite{note3}  
That is, any fermion wavefunction  
which respects the following local constraints is an element 
of our Hilbert space: 
\begin{eqnarray} 
\Big\{\sum_{a=1}^{N}  
{\rm Tr}\big[
\Psi^{a\dagger}_{j}\sigma_{\mu}
\Psi^{a}_{j}\big] \Big\} \big|{\rm phy} \big\rangle \equiv 0,  
\hspace{0.3cm} \forall j,\mu=1,2,3 \ . \nn  
\end{eqnarray}  
The density and spin operators defined in Eq.~(\ref{s+-}) 
in fact act within this physical Hilbert space. Moreover, 
they observe the following commutation relations:~\cite{note3}  
\begin{align}
\big[S^{ab3},S^{cd3}\big] &= \frac{1}{2}
\big( \delta^{bc}S^{ad3} - \delta^{ad}S^{cb3}\big), \nn \\
\big[S^{ab3},S^{cd+}\big] &= \frac{1}{2}
\big( \delta^{bc}S^{ad+} + \delta^{bd}S^{ac+}\big), \nn \\ 
\big[S^{ab3},S^{cd-}\big] &= - \frac{1}{2}
\big( \delta^{ad}S^{bc-} + \delta^{ac}S^{bd-}\big), \nn \\ 
\big[S^{ab+},S^{cd+}\big] & = 0, \ \ 
\big[S^{ab-},S^{cd-}\big] = 0, \nn \\
\big[S^{ab+},S^{cd-}\big] &= \frac{1}{2}
\big( \delta^{ac}S^{bd3} + \delta^{ad}S^{bc3} 
+ \delta^{bc}S^{ad3} + \delta^{bd}S^{ac3} \big), \nn \\ 
\big[\psi^{ab},S^{cd\pm}\big] &= \frac{i}{2}
\big( \delta^{bc}S^{ad\pm} - \delta^{ac}S^{bd\pm} 
- \delta^{ad}S^{cb\pm} + \delta^{bd}S^{ca\pm} \big), \nn \\  
\big[\psi^{ab},S^{cd3}\big] &= \frac{i}{2}
\big( \delta^{bc}S^{ad3} - \delta^{ac}S^{bd3} 
- \delta^{ad}S^{cb3} + \delta^{bd}S^{ca3} \big), \nn \\ 
\big[\psi^{ab},\psi^{cd}\big] &= \frac{i}{2}
\big( \delta^{bc}\psi^{ad} - \delta^{ac}\psi^{bd} 
- \delta^{ad}\psi^{cb} + \delta^{bd}\psi^{ca} \big). \label{algebra} 
\end{align} 
Using them, one can argue that  
the generalized spin Hamiltonian given in Eq.~(\ref{largeN}) 
is invariant under those continuous symmetries which  
are generated by  
\begin{eqnarray}
\psi^{ab}_{\rm tot}, \ \ \sum^N_{a=1} S^{aa3}_{\rm tot}, \ \ 
\sum^N_{a=1} S^{aa1}_{\rm tot}, \ \ 
\sum^N_{a=1}  S^{aa2}_{\rm tot}. \nn 
\end{eqnarray}       
When $N=1$, $\psi^{ab}$ disappears by itself and 
Eq.~(\ref{largeN}) in combination with Eq.~(\ref{algebra}) 
reduces to the $SU(2)$ Heisenberg spin model defined in 
Eq.~(\ref{ff}).


\begin{thebibliography}{99}
\bibitem{note} Even in the presence of time-reversal symmetry, 
the mutual (double) Chern-Simon mechanism sometimes protects a given  
mean-field ansatz from massless gauge fluctuations\cite{diamantini}. 
However, we expect that it does not work either here, since our 
BW state belongs to the `weak topological 
insulator' -- ${\cal Z}_2$ even class --  
(see sec.~IV).
\end{thebibliography}
\end{document}